\documentclass[10pt,number]{elsarticle}

\usepackage{graphicx}
\usepackage{epstopdf}
\usepackage{dcolumn}
\usepackage{bm}

\begin{document}
\begin{frontmatter}

\title{Rejection-free kinetic Monte Carlo simulation of multivalent biomolecular interactions}

\author[label 1]{Jin Yang\corref{cor1}}
\author[label 2,label 3]{William S. Hlavacek}

\address[label 1]{Chinese Academy of Sciences -- Max Planck Society Partner
Institute for Computational Biology, Shanghai Institutes for Biological
Sciences, Shanghai 200031, China}  
\cortext[cor1]{Correspondence should be addressed to J.Y. (E-Mail:
yangjin@picb.ac.cn, Tel: +86-21-5492-0476, Fax: +86-21-5492-0451)} 

\address[label 2]{Theoretical Division and Center for Nonlinear Studies, Los
Alamos National Laboratory, Los Alamos, NM 87545, USA} 
\address[label 3]{Department of Biology, University of New Mexico, Albuquerque, NM 87131, USA} 

\begin{abstract}
\small{The system-level dynamics of multivalent biomolecular interactions can be simulated using a rule-based kinetic Monte Carlo method in which a rejection sampling strategy is used to generate reaction events.  This method becomes inefficient when simulating aggregation processes with large biomolecular complexes.  Here, we present a rejection-free method for determining the kinetics of multivalent  biomolecular interactions, and we apply the method to simulate simple models for ligand-receptor interactions. Simulation results show that performance of the rejection-free method is equal to or better than that of the rejection method over wide parameter ranges, and the rejection-free method is more efficient for simulating systems in which aggregation is extensive. The rejection-free method reported here should be useful for simulating a variety of systems in which multisite molecular interactions yield large molecular aggregates.}
\end{abstract}

\end{frontmatter}

\section{Introduction}
Protein-protein interactions in signal transduction involve domain-based protein interactions and site-specific modifications~\cite{hunter2000signaling,Pawson2003}. Simulating the dynamics of a complex signaling system that has many protein interactions is usually a daunting task because a large (bio)chemical reaction network is typically required to model interactions at the level of site-specific details and submolecular domains~\cite{HlavacekWS:TheCCS,Hlavacek07182006,danos2007rbm,mayer2009molecular}. Even though a large-scale biochemical reaction network can be built by either manual or automated construction~\cite{shapiro2003cellerator,Blinov11222004,Lok2005NatB,mallavarapu2009programming}, simulating such models is computationally inefficient because a conventional kinetic Monte Carlo simulation algorithm, for example, has a cost that depends on the size of a network measured by the number of reactions~\cite{gillespie2007ssc} or the number of chemical species~\cite{ramaswamy2009new}.

\begin{figure}
\centering
\includegraphics[scale=0.6]{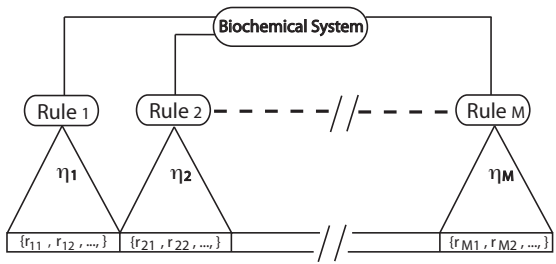}
\caption{\label{fig:rbm} Diagrammatic depiction of a biochemical system described by rules
and its underlying reaction network. Rules partition the entire reaction list
into disjoint subsets which are consolidated by rules into rate processes
denoted by $\{\eta_1,\eta_2,...,\eta_M\}$. The term $r_{ij}$ denotes the $j$th reaction
implied by Rule $i$. Note that partitions of a reaction network specified by
rules are not necessarily of finite size, which is illustrated in our model of multivalent ligand-receptor interactions (see main text). Rule-based kinetic Monte Carlo samples the rule list (the intermediate layer) to generate reaction events and avoids simulating a system by means of a reaction network (the bottom layer).} 
\end{figure}

\begin{figure}
   \begin{center}      
      \includegraphics{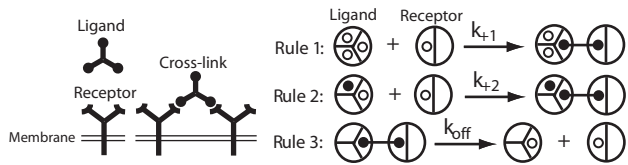}
      \caption{The interactions of a trivalent ligand and a bivalent
cell-surface receptor (left). Graphical rules (right) that represent free ligand recruitment to the cell surface (Rule 1), receptor crosslinking by ligand (Rule 2) and ligand-receptor bond dissociation (Rule 3). Rules are parameterized by rate constants $k_{+1}$, $k_{+2}$ and $k_{\rm off}$, respectively. An empty (filled) circle inside a container denotes a free (bound) site. A connecting line indicates a ligand-receptor bond. An empty container indicates an arbitrary state (free or bound). An additional assumption is imposed to prohibit reactions defined by $\rm R_2$ from forming cyclic ligand-receptor aggregates (see main text).}
   \label{fig:transtree}
   \end{center}
\end{figure}

The challenge of simulating protein-protein interactions in signal transduction can be addressed with a rule-based modeling paradigm (see Ref.~\cite{Hlavacek07182006} for a review). Rule-based modeling provides a hierarchical structure to define biochemical reaction systems (Fig.~\ref{fig:rbm}). In a rule-based approach, molecules are modeled as structured objects composed of reactive sites, and reaction rules are used to represent
interactions~\cite{Hlavacek07182006,faeder2009rule,feret2009internal} (see
Fig.~\ref{fig:transtree} for examples of rules for ligand-receptor interactions).  In general, a rule specifies local properties of individual sites (e.g., whether a site is free or occupied) in a molecule and application conditions that require checking non-local information (e.g., whether two sites are members of the same macromolecular aggregate). Assuming rate laws for elementary reactions, one parameterizes the reaction classes implied by a rule with a specific rate constant. Thus, a rule provides a compact representation of these reactions based on the law of mass action. 

Kinetic Monte Carlo (KMC) methods have been developed for simulating the stochastic dynamics of rule-based models~\cite{Danos2007,Yang_Arxiv07,colvin2009simulation}. The methods of Danos et al.~\cite{Danos2007} and Yang et al.~\cite{Yang_Arxiv07} avoid the requirement of specifying a chemical reaction network prior to simulation by directly sampling a rule list to generate reaction events and updating the system state in accordance with reaction rules. The computational cost is essentially independent of network size because a rule list is typically more compact and orders of magnitude smaller than the size of the corresponding reaction network.

The rule-based KMC simulation method of Yang et al.~\cite{Yang_Arxiv07} involves rejection sampling. After a rule is selected, trial sites with suitable local properties are randomly sampled to potentially undergo the chemical transformation(s) specified by the rule. Trial sites are rejected if they are not compatible with the application conditions of the rule. Random site selection in the rejection sampling steps is a constant-time operation, but efficiency might be significantly compromised in cases where null events become a dominant fraction of all sampled events. This scenario happens when the method is used to simulate multivalent ligand-receptor interactions in a sol-gel region that yields a giant ligand-receptor aggregate at equilibrium, where sampled trial sites are very likely rejected because the vast majority of sites reside in the giant aggregate and are prohibited from binding to each other by an application condition (a model assumption) that prohibits intra-aggregate binding reactions to form cyclic aggregates.

In this work, we present a rejection-free KMC method for simulating the kinetics of reactions implied by rules. This approach can be used to efficiently simulate processes involving multivalent biomolecule clustering and aggregation, processes often encountered in biochemical systems. The method samples sites for reaction according to their exact probability of taking part in a reaction defined by a sampled rule and consequently avoids rejecting trial sites altogether. The simulation is statistically exact and is validated by comparing to continuum solutions of low-dimensional models for multivalent ligand-receptor binding system. We simulate a trivalent ligand -- bivalent receptor model using the method of Yang et al.~\cite{Yang_Arxiv07} and the rejection-free method. The latter method is found to have the same or better efficiency for most parameter values. 

\section{Rule-based kinetic Monte Carlo}
In this section, we describe the strategy of rejection-free KMC for rule-based models. In a biomolecular reaction system, particularly in a system involving protein-protein  interactions for signal transduction, proteins reversibly bind to one another via their binding sites (conserved domains). Protein sites (amino acid residues) can also be modified, by phosphorylation, methylation, etc. The state of the system (or a system configuration) is determined by states of individual sites and their connectivities. The temporal dynamics of a biochemical reaction system is the evolution of the system state in time, which can be modeled as a Markov process that is generally described by the Chapman-Kolmogorov equation (or more specifically, the chemical master equation for our case)~\cite{van2007stochastic}: 
\begin{equation}\label{eq:cke}
\frac{dP(Y,t)}{dt}= \sum_{Y'\ne Y} \left[W(Y|Y')P(Y',t) - P(Y,t)W(Y'|Y) \right] \ .
\end{equation}
$P(Y,t)$ is the probability that the system is found in state $Y$, and $W(Y'|Y)$ gives the transition rate from state $Y$ to state $Y'$. In a conventional chemical reaction system, a state $Y$ is defined by the concentrations of all chemical species. In a rule-based system, a state $Y$ is defined by the site occupancy and connectivity of individual molecules. Analytical solutions to the above master equation are only possible for very simple systems. Direct numerical integration of the above ordinary differential equations is also intractable because the size of the state space is often enormous even for a mildly-complex system that involves interactions among a few protein types. To solve this problem, Monte Carlo simulation is often applied to conduct random walks over the state space and to generate stochastic state trajectories for a system of interest. 

Here we consider a biochemical system that is described by a set of reaction rules (see Fig.~\ref{fig:rbm}), ${\mathbf
R}=\{\rm R_1,...,R_M\}$. The system evolves in time with rates
{\boldmath$\mathbf{\eta}$}$=\{\rm \eta_1,...,\eta_M\}$ for the rules. Such rate processes can be simulated using the standard kinetic Monte Carlo method~\cite{voter2007introduction}. The waiting time $\tau$ between two consecutive (reaction) events is taken to follow an exponential distribution
\begin{equation}\label{eq:taup}
P(\tau)=\eta_{\rm tot} e^{-\eta_{\rm tot}\tau} \ , 
\end{equation}
where $\eta_{\rm tot}=\sum_{i=1}^M\eta_i$ is the
sum of the rule rates. The mean waiting time for a reaction event is the inverse of $\eta_{\rm tot}$, $\langle\tau\rangle=\frac{1}{\eta_{\rm tot}}$. For each reaction event, a waiting time $\tau$ is sampled using the following equation:
\begin{equation}\label{eq:tau}
\tau=-\frac{\ln(\rho_1)}{\eta_{\rm tot}} \ ,
\end{equation}
where $\rho_1$ is a uniform random number in $(0,1)$. The probability for a rule to be chosen is proportional to its rate. Therefore, a specific rule $e$ is sampled by finding the integer $e$ that satisfies: 
\begin{eqnarray}\label{eq:e}
\sum_{i=1}^{e-1}\eta_i <\rho_2\eta_{\rm tot}\le \sum_{i=1}^e\eta_i \ ,
\end{eqnarray}
where $\rho_2$ is a uniform random number in $(0,1)$. 

The rule-based kinetic Monte Carlo algorithm requires several additional operations per reaction event following the selections of $\tau$ and $e$: (1) select reactive sites admissible to the rule; (2) excute the chemical transformation defined by rule $e$; and (3) update components in the rate vector {\boldmath$\mathbf\eta$} that are affected by the chemical transformation.

To choose reactive sites, we can write the rule-specific probability for a site $x$ in a candidate set $X_e$ to be chosen as (we note that a ``site" $x$ denotes a set of interacting protein sites that are defined by the rule. For example, for a bimolecular interaction, $x$ will include two interacting sites)
\begin{equation}
 P(x)=P_{a|s}(x)P_s(x) \ .
\end{equation}
This equation describes a two-step procedure to find a site $x$: (a) choose a site  $x$  according to the sampling probability $P_s(x)$ and (b) accept the chosen $x$ according to the acceptance probability $P_{a|s}(x)$. In the rejection method, a convenient sampling distribution $P_s(x)$ of candidate sites, usually a uniform distribution for maximum sampling efficiency, is used to generate trial sites and trial sites may be rejected according to the acceptance probability $P_{a|s}(x)$ so that the true probability $P(x)$ is recovered. The rejection sampling introduces a ``null" process by sampling nonreactive sites such that this null process advances time but does not change the system state. Therefore, we can partition the rate of a rule into two components:
\begin{equation}\label{eq:eta}
\eta_e=\eta_{e,a}+\eta_{e,n} \ \ \ e=1,...,M \ ,
\end{equation}
where $\eta_{e,a}$ and $\eta_{e,n}$ denote the actual reactive rate and the null process rate of rule $e$, respectively. Generally, the size of the null fraction of $\eta_{e,n}$ depends on the site sampling procedure which also affects the calculation of the rule rate. Operating with the combined rate $\eta_e$ ignores any application condition by model assumption imposed on this rule in the step of site sampling. The model assumption can however be enforced by a subsequent rejection of the sampled sites that are non-permissible to react. Because the rejection method often can achieve random sampling of reactive sites, the cost per sampling step is constant and the implementation is usually straightforward. Determining whether to accept trial sites for reaction may however incur non-constant cost and complex bookkeeping, which depends on the nature of model assumptions.

The effect of rejection sampling on computational performance can be measured by the rejection ratio:
\begin{equation}\label{eq:theta}
\theta=\frac{\eta_{\rm tot,n}}{\eta_{\rm tot}} \ ,
\end{equation}
where $\eta_{\rm tot,n}=\sum_{i=1}^M{\eta_{i,n}}$ is the total rate of null
processes. For some systems where $\theta$ may be close to 1, a majority of
trial sites will be rejected and the algorithm becomes inefficient. For such
situations, an algorithm without rejections (or with a substantially reduced
number of rejections) is desirable.

In contrast, a rejection-free method always accepts sampled site $x$. In other words, $P_{a|s}(x)$ is unity for every $x$ and $P(x)$ is captured at the step of site sampling, i.e., $P(x)=P_s(x)$. 

An early rejection-free Monte Carlo simulation method was developed by Bortz et al.~\cite{bortz1975new} to simulate Ising spin systems, which overcomes the significant slowing down near the critical temperature by the classic Metropolis method~\cite{metropolis1953equation}. The algorithm is often called the BLK algorithm, or the $n$-fold way method, and it has been widely used to simulate systems with high density and/or at low temperature. Gillespie later developed a stochastic simulation algorithm, equivalent to the method of Bortz et al.~\cite{bortz1975new}, for simulating coupled chemical reactions in a homogeneous reaction compartment~\cite{gillespie1977exact}. Below, we describe a rejection-free procedure for simulating rule-based models.

Without loss of generality, we only consider unimolecular and bimolecular reactions. Higher-order reactions can usually be decomposed into these two types of elementary reactions. For a unimolecular reaction (e.g., a single-step site modification or a noncovalent bond dissociation between two proteins) of rule $e$ that involves a single type of protein site $A$, at a given time, the candidate sites are defined in a set $X_e=\{a_i|i=1,...,n_a\}$ that includes all legitimate sites designated by the rule and its application conditions. In many cases, by the mean-field approximation, sites available according to a rule definition are considered to have equal probability to be chosen. The probability of a site $x=a_i$ to be chosen is 
\begin{equation}\label{eq:ai}
 P(x=a_i) = P_s(x=a_i) = \frac{w_i}{Z} \,
\end{equation}
where $w_i$ is the reaction propensity for site $a_i$ and $Z$ is a partition function $Z=\eta_e=\sum_{X_e}w_i$. In general $w_i$ reflects the biophysical property that determines the reaction propensity of the site $a_i$. In most applications, a rule usually defines reactions of sites from the same protein types and with identical properties. For this reason and under the law of mass action, we can interpret $w_i$ as the specific rate (or, the rate constant) $k_e$ for the rule. The partition funtion becomes the total rule rate, $Z=k_e|X_e|$, where $|X_e|$ denotes the size of the candidate set $X_e$ (i.e., $|X_e|=n_a$). Effectively, Eq.~\ref{eq:ai} defines a uniform sampling distribution over $X_e$. The above analysis can be readily extended to the case of bimolecular reactions. Considering an interaction such as binding between two types of protein sites $A$ and $B$, we have a candidate set $X_e=\{a_i,b_j|i=1,...,n_a, j=1,...,n_b\}$. The probability of a pair of sites $x=(a_i,b_j)$ to be chosen to react is 
\begin{equation}
 P(x=\{a_i,b_j\})=P_s(x=\{a_i,b_j\})=\frac{w_{ij}}{Z} \ .
\end{equation}
Similarly, $w_{ij}$ can be interpreted as a specific rate $k_e$ that is identical for all site pairs. According to the law of mass action, the partition function $Z=\sum_{X_e}w_{ij}=k_e|X_e|=k_en_an_b$.

Based on the above considerations, we summarize the rejection-free algorithm as follows.

\begin{enumerate}
\item Initialization: assign copy numbers of proteins and specify initial states of individual protein sites, assign rate constants {\boldmath{$k$}} for rules, and calculate the initial values of {\boldmath{$\eta$}}. 
\item Sample a waiting time $\tau$ and choose a reaction rule $e$ according to Eqs.~(\ref{eq:tau}) and~(\ref{eq:e}), respectively. 
\item Sample protein sites based on the distribution $P(x)$,
update states of the chosen sites according to the rule specification, and 
recalculate rate vector \boldmath{$\eta$}. 
\item Repeat Steps 2 and 3.
\end{enumerate}

The above steps provide a general strategy for simulating rule-based biochemical reaction systems. In the following section, we apply the method to simulate a multivalent ligand-receptor interaction model and use the model to discuss the implementation details.

\section{Multivalent ligand-receptor binding systems}

Cell-surface receptor aggregation induced by ligand binding is an important step in many signal transduction pathways~\cite{goldstein2004mac}. In an earlier study, Goldstein and Perelson developed an equilibrium model to investigate binding interactions between trivalent extracellular ligands and bivalent cell-surface receptors~\cite{BGoldstein06011984}. The Goldstein-Perelson model predicted in theory that for certain parameter values small receptor aggregates and  ``superaggregates'' can coexist at equilibrium. In fact, sol-gel phase transition phenomena can happen in any multivalent ligand-receptor system in which either the ligand or the receptor has more than two binding sites and each ligand and receptor has at least two binding sites. 

Recently, Yang et al.~\cite{Yang_Arxiv07} developed a rule-based kinetic Monte Carlo method that was used to simulate the stochastic dynamics of a kinetic version of the Goldstein-Perelson model, which has been called the TLBR model, where TLBR represents ``trivalent ligand -- bivalent receptor." The gelation processes of ligand-receptor aggregation were observed in simulations, which was consistent with the Goldstein-Perelson theory of sol-gel phase transition. The simulation further revealed the kinetic details about formation of superaggregates. Monine et al.~\cite{monine2009bj} also applied the same method to study steric effects using extended kinetic versions of the Goldstein-Perelson model. The KMC simulation method of Yang et al.~\cite{Yang_Arxiv07} includes rejection sampling, which becomes inefficient when a system forms a superaggregate that contains most ligands and receptors. In this section, we evaluate the rejection-free method approach by applying it study multivalent ligand-receptor interaction systems.

\subsection{Model of ligand-receptor binding}
We consider a system with $N_L$ ligands and $N_R$ receptors. The ligand and receptor have $v_l$ and $v_r$ symmetric binding sites, respectively.  Three types of reactions are considered:  (1) free ligand recruitment to a receptor on the cell surface, (2) crosslinking of two cell-surface receptors by a ligand that is already bound on the cell surface but has at least one free site, and (3) dissociation of a ligand-receptor bond.  The rule-based specification of the TLBR model, a kinetic extension of the Goldstein-Perelson model, is illustrated in Fig.~\ref{fig:transtree}. The model has three reaction rules (Rules 1-3). To compare results with the Goldstein-Perelson model, we impose the same model assumptions: a) binding sites are equivalent~\cite{Perelson1980}, b) a ligand cannot associate with a receptor via more than one bond, and c) a ligand cannot associate with a receptor that is a member of the same aggregate (an aggregate is a ligand-receptor complex that has at least one ligand and one receptor in it). These assumptions prevent the formation of cyclic ligand-receptor aggregates and affect the calculation of rule rates, which is the actual reason that distinguishes rejection and rejection-free sampling prcedures.

\subsection{Calculation of rule rates and site sampling}
All rule rates are calculated according to the law of mass action. We first give equations for directly calculating $\eta_1$ and $\eta_3$. Calculation of $\eta_2$ requires non-trivial treatment, and different treatment implies different implementation of the algorithm. We will explain these issues in detail.

The rate of free ligand recruitment to a cell-surface receptor for Rule 1, $\eta_1$, is proportional to the product of the number of free sites on free ligands and the number of free receptor sites on the cell surface, 
\begin{equation}
\eta_1=k_{+1}v_lF_L(v_rN_R-B) \ , 
\end{equation}
where $F_L$ is the number of free ligands. We note that Rule 1 specifies a class
of bimolecular reactions and the rate constant $k_{+1}$ absorbs a volume factor (this also applies to the rate constant $k_{+2}$). The rate of ligand-receptor bond dissociation  for Rule 3, $\eta_3$, is proportional to the number of ligand-receptor bonds,  
\begin{equation}
\eta_3=k_{\rm off}B \ .
\end{equation}
The number of bonds $B$ increases (decreases) by 1 upon an association (dissociation) event.

\begin{figure}
\begin{center}
\hspace{0.2in}\includegraphics[scale=0.3]{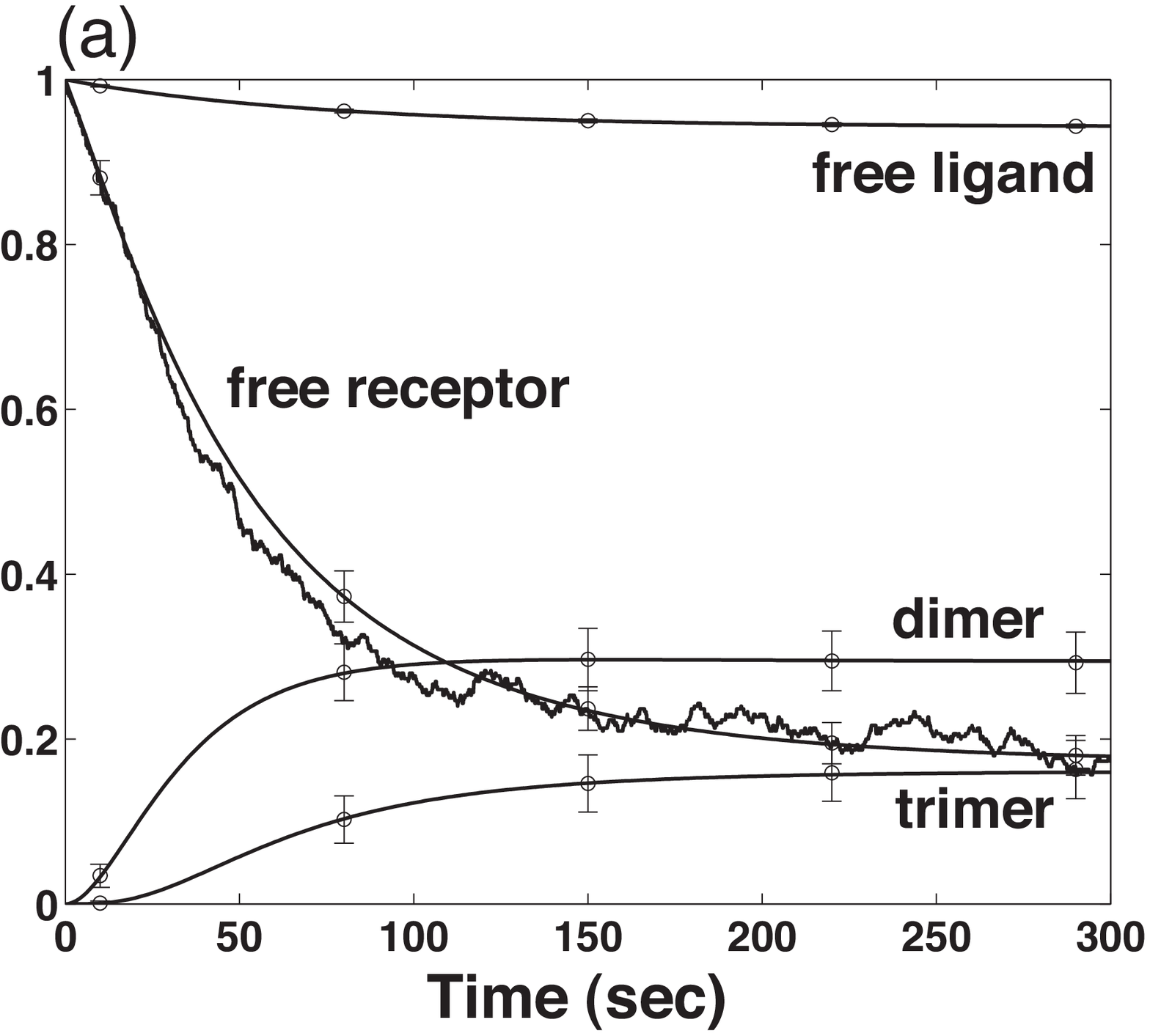}
\includegraphics[scale=0.3]{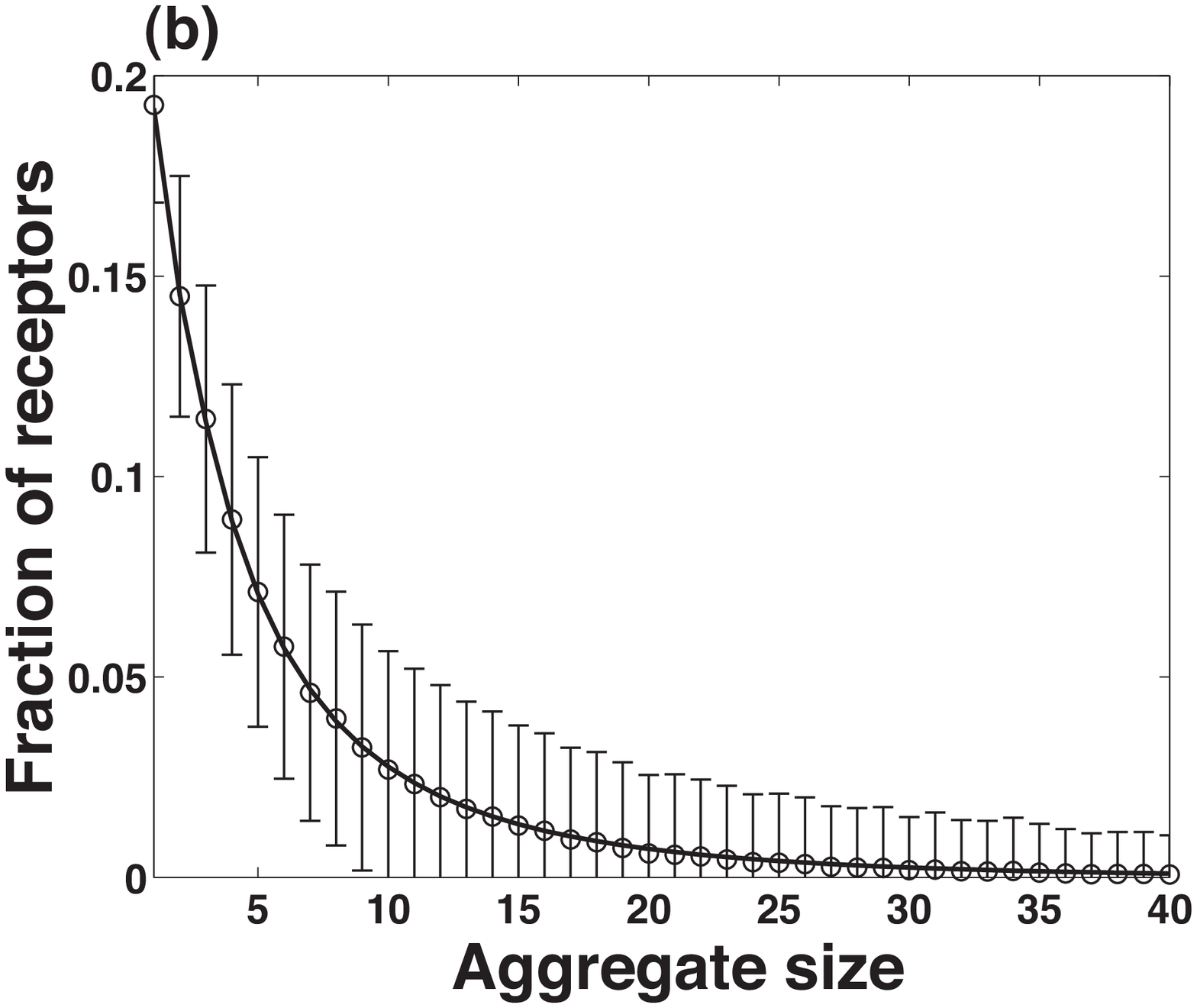}
\caption{Validation of the simulation algorithm for multivalent ligand-receptor interaction systems. (a) Comparison of normalized results to the ODE solutions for the bivalent ligand and bivalent receptor interaction system. The normalization factor is $N_R/n$, where $n$ is the number of receptors in an aggregate ($n=2$ for dimers, $n=3$ for trimers, etc.). The means and standard deviations of the simulation results obtained by the simulation method are shown on top of the continuous ODE solutions. One stochastic time trajectory is shown for the free receptor population. (b) Stochastic receptor aggregate distribution for the trivalent ligand and bivalent receptor system. The system reaches equilibrium after 350 seconds and the averages of equilibrium distributions match the results (solid curve) obtained using the model of Goldstein and Perelson~\cite{BGoldstein06011984}. Parameter values: $N_R=300$, 
$N_L=4200$, $k_{+1}=6.67\times 10^{-7}$ s$^{-1}$, $k_{+2}=100k_{+1}$, $k_{\rm off}=0.01$
s$^{-1}$. Initial condition: all simulations start with free ligands and free receptors without bonds.} 
      \label{fig:vld}
   \end{center}
\end{figure}

For ligand-mediated receptor cross-linking on the cell surface (an event of Rule 2), to avoid forming cyclic ligand-receptor aggregates, one must ensure that an association event joins either two separate ligand-receptor aggregates or a ligand-receptor aggregate and a free receptor.  We note that the probability for a ligand site to be chosen is proportional to the number of free receptor sites to which  the ligand site may bind, i.e., the number of free receptor sites excluding ones in the same aggregate with the candidate ligand site. Once a ligand site is chosen, the probability of selecting a binding receptor site is uniform for all free receptor sites that are not in the same aggregate with the selected ligand site.

A straightforward approach is to calculate the probability for each free receptor site to take part in the next event and the probability for each free ligand site (on the cell surface) to crosslink the free receptor site. However, the cost of searching over sites and updating site probabilities would scale with the number of molecules per Monte Carlo step.

At a given time, if the system has $N_A$ ligand-receptor aggregates on the cell surface (excluding free receptors), the rate $\eta_2$ can be given by a direct sum over combinations between distinct aggregate pairs, i.e.,  
\begin{equation} \label{eq:eta2}
\eta_2=k_{+2}\left[v_rF_R(v_l(N_L-F_L)-N_B)+\sum_{i=1}^{N_A}\sum_{j=1, j\ne
i}^{N_A}\left(l_ir_j+l_jr_i\right)\right ] \ ,
\end{equation}
where $l_i$ and $r_i$ are the numbers of free ligand and receptor sites in the $i$th aggregate, respectively, and $F_R$ is the number of free receptors. The term $v_rF_R(v_l(N_L-F_L)-N_B)$ accounts for the interactions between free ligand sites on the cell surface and free receptor sites. However, the computational cost of Eq.~(\ref{eq:eta2}) is $O(N_A^2)$, which is impractical when $N_A$ is large. Alternatively, rate $\eta_2$ can be calculated as
\begin{equation}\label{eq:u2}
\eta_2 = k_{+2}\sum_{i=1}^{N_A} l_i(v_rN_R-B-r_i) \ ,
\end{equation}
The term $(v_rN_R-B-r_i)$ is the total number of free receptor sites that are not in aggregate $i$. From the above equation, it follows that the probability that an aggregate $i$ contributes a ligand site for a Rule 2 event is 
\begin{equation}\label{eq:pi}
P_i = \frac{k_{+2}l_i(v_rN_R-B-r_i)}{\eta_2}, \ \ \ i=1,...,N_A \ .
\end{equation}
We first sample this distribution to locate an aggregate $i$ that provides a free ligand site. All free ligand sites in aggregate $i$ are equivalent (according to model assumptions) and therefore each of them has an equal probability to be chosen. Then, a free receptor site that is not in aggregate $i$ is chosen to bind with the sampled ligand site. Similarly, all such free receptor sites (either in another aggregate or in the free receptor pool) are equivalent and therefore can be randomly sampled. The number of free ligand and receptor sites, $l_i$ and $r_i$, can be calculated when the numbers of ligands and receptors in an aggregate are known. In an acyclic aggregate that has $n_r$ receptors and $n_l$ ligands, the number of bonds is given as $b=n_r+n_l-1$, and the numbers of free receptor and ligand sites can be calculated as $r=v_ln_l-b$ and $l=v_rn_r-b$, respectively. This corresponds to a graph-theoretic observation that a connected acyclic graph (equivalent to a tree) has a number of edges one less than the number of nodes~\cite{bollobas1998modern}. A ligand-receptor aggregate can be represented as a bipartite graph, in which receptors and ligands are nodes and ligand-receptor bonds are edges. The numbers of receptors and ligands can be obtained by a linear-time traversal of the aggregate graph. A better, alternative approach is to maintain an aggregate list during simulation. This approach avoids routine graph traversal of aggregate graphs for each reaction event except for bond dissociation events, in which graph traversals are required to determine the resulting two separate aggregates.  Direct calculation of $\eta_2$ using Eq.~(\ref{eq:u2}) has a linear-time cost scaled by the number of aggregates $N_A$. To achieve a constant-time cost per event, we update $\eta_2$  iteratively by accounting for the change caused by a reaction event. Eq.~(\ref{eq:u2}) can be rewritten as follows:
\begin{equation}\label{eq:u2a}
\eta_2=k_{+2}\left([v_l(N_L-F_L)-B](v_rN_R-B)-\sum_{i=1}^{N_A}l_ir_i\right) \ .
\end{equation}
The term $[v_l(N_L-F_L)-B](v_rN_R-B)$ accounts for all pair combinations of free ligand sites (on the cell surface) and free receptor sites under the law of mass action. We note that, in a method using rejection sampling~\cite{Yang_Arxiv07}, this quantity is used to calculate rate $\eta_2$ and a trial pair of ligand and receptor sites can be randomly chosen but are subject to rejection later if they are found to be components of the same aggregate. For our rejection-free implementation, the exact rate is obtained by subtracting $u\equiv\sum_{i=1}^{N_A}l_ir_i$, where $u$ is the number of intra-aggregate site-pair combinations, to accommodate the model assumption that prohibits intra-aggregate binding reactions. The first term in Eq.~(\ref{eq:u2a}) can be easily recalculated with a constant cost after each reaction event. The second term can be updated iteratively by $\Delta u$ (Table~\ref{tab:du2}). Each reaction event changes the list of aggregates by creating a new one, or modifying or removing existing ones. For instance, a free ligand binding to a receptor may (if the receptor is also free) or may not (if the receptor is already bound to another ligand) create a new aggregate. However, most aggregates do not participate in a reaction event and thus remain unchanged. Therefore, updating Eq.~(\ref{eq:u2a}) is straightforward by accounting for sites on a small number of affected aggregates (usually one or two) and it has a constant-time cost. For example, a merge of two aggregates $i$ and $j$ creates one new aggregate $k$ ($N_A$ decreases by 1). In this case, on the newly-formed aggregate, free ligand and receptor sites both decrease by 1 from the sums of those on the two merging aggregates, and it follows that $\Delta u=l_kr_k-l_ir_i-l_jr_j$, where $l_k=l_i+l_j-1$ and $r_k=r_i+r_j-1$. Formulas for all event types are given in Table~\ref{tab:du2}.

\begin{table}
\caption{\label{tab:du2} Formulas for calculating  $\Delta u_2$}
\footnotesize
\begin{tabular}{cll}\hline
Rule & Event & $\Delta u_2$ \\\hline\hline
1 & A free ligand binds to a free receptor & $(v_r-1)(v_l-1)$ \\
& A free ligand binds to aggregate $i$ & $(v_l-1)(r_i-1)-l_i $ \\\hline
2 & Aggregate $i$ associates with a free receptor & $(v_r-1)(l_i-1)-r_i$ \\
& Aggregates $i$ and $j$ associate to form a new aggregate $k$ & $l_kr_k-l_ir_i-l_jr_j$ \\\hline
3 & An aggregate breaks into a free ligand and a  free receptor & $-(v_r-1)(v_l-1)$ \\
& A ligand dissociates from aggregate $i$ &$l_i-(v_l-1)(r_i+1)$ \\
& A receptor dissociates from aggregate $i$ & $r_i-(v_r-1)(l_i+1)$ \\
&  Aggregate $k$ dissociates into two aggregates $i$ and $j$ & $l_ir_i+l_jr_j-l_kr_k$ \\\hline
\end{tabular}
\end{table}

Our implementations of the rejection method of Yang et al.~\cite{Yang_Arxiv07} and the rejection-free method reported here are available upon request. These implementations simulate the TLBR model, as well as models for $m$-valent ligand and $n$-valent receptor that incorporate model assumptions similar to those of the Goldstein-Perelson model. The rejection and rejection-free codes were written with the intention of making them as much alike as possible, except for the sampling strategies and rule rate calculations,  to eliminate irrelevant differences. We note that our implementations of the simulation methods discussed here provide new tools for studying the equilibria and kinetics of multivalent ligand-receptor interactions and gelation on cell membranes, topics which have been studied intensely over the years via a variety of approaches~\cite{perelson1984some,lauffenburger1996receptors,guo1999thermodynamic, george2003aggregation}.

\section{Simulation results}
In this section, we present results obtained using the rejection-free algorithm described above. Parameter values used in simulations were taken from Ref.~\cite{Yang_Arxiv07}. 

To validate the accuracy of our method, we compare simulation results with those
obtained using ordinary differential equations
(ODEs). The relaxation kinetics of a bivalent ligand ($v_l=2$) -- bivalent
receptor ($v_r=2$) system to equilibrium is shown in Fig.~\ref{fig:vld}(a). The
deterministic results were obtained by solving a small number of ODEs as
described in Refs.~\cite{Perelson1980,Posner1995}. The stochastic trajectories
recapitulate on average the deterministic solutions. In Fig.~\ref{fig:vld}(b),
the equilibrium distribution of receptor aggregates in a TLBR system agrees on average with the equilibrium results from the Goldstein-Perelson model. These results demonstrate that the simulation algorithm produces outcomes consistent with those obtained independently. 

A phase transition in ligand-receptor clustering at equilibrium is predicted in 
the Goldstein-Perelson model by varying two dimensionless parameters $c_{\rm tot}=3k_{+1}N_L/k_{\rm off}$ and $\beta=k_{+2}N_R/k_{\rm off}$, which shows that finite-sized ligand-receptor aggregates can coexist with an infinite-sized polymer-like aggregate in a so-called sol-gel coexistence regime~\cite{BGoldstein06011984}. To investigate the
stochastic dynamics of ligand-receptor aggregation and the algorithmic performance at different phase regimes, we use the average receptor aggregate size to measure the degree of ligand-receptor clustering, which can be calculated as  

\begin{figure}
\centering
\includegraphics[scale=0.3]{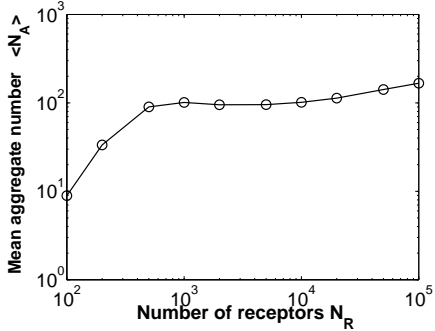}
\caption{\label{fig:nrnc} The relationship between $\langle N_A\rangle$ and $N_R$ for the trivalent ligand--bivalent receptor system. The number of ligands $N_L$ is set equal to the number of receptors $N_R$. $\langle N_A\rangle$ is obtained by calculating its mean value from $10^6$ consecutive reaction events after the system reaches equilibrium. Other parameters and initial conditions are identical to the values indicated in Fig.~\ref{fig:vld}.} 
\end{figure}

\begin{figure}
\centering
\includegraphics[scale=0.3]{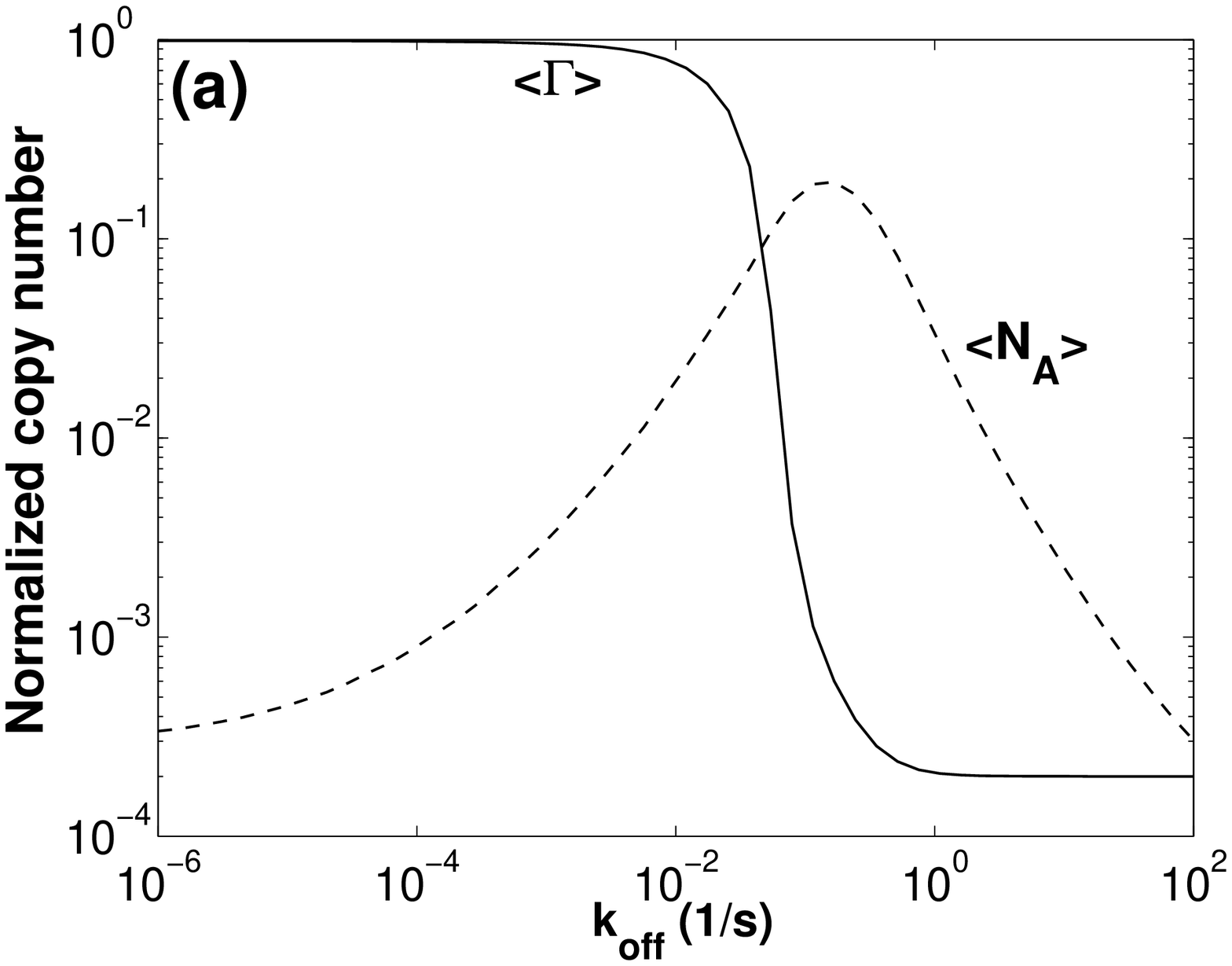}

\includegraphics[scale=0.3]{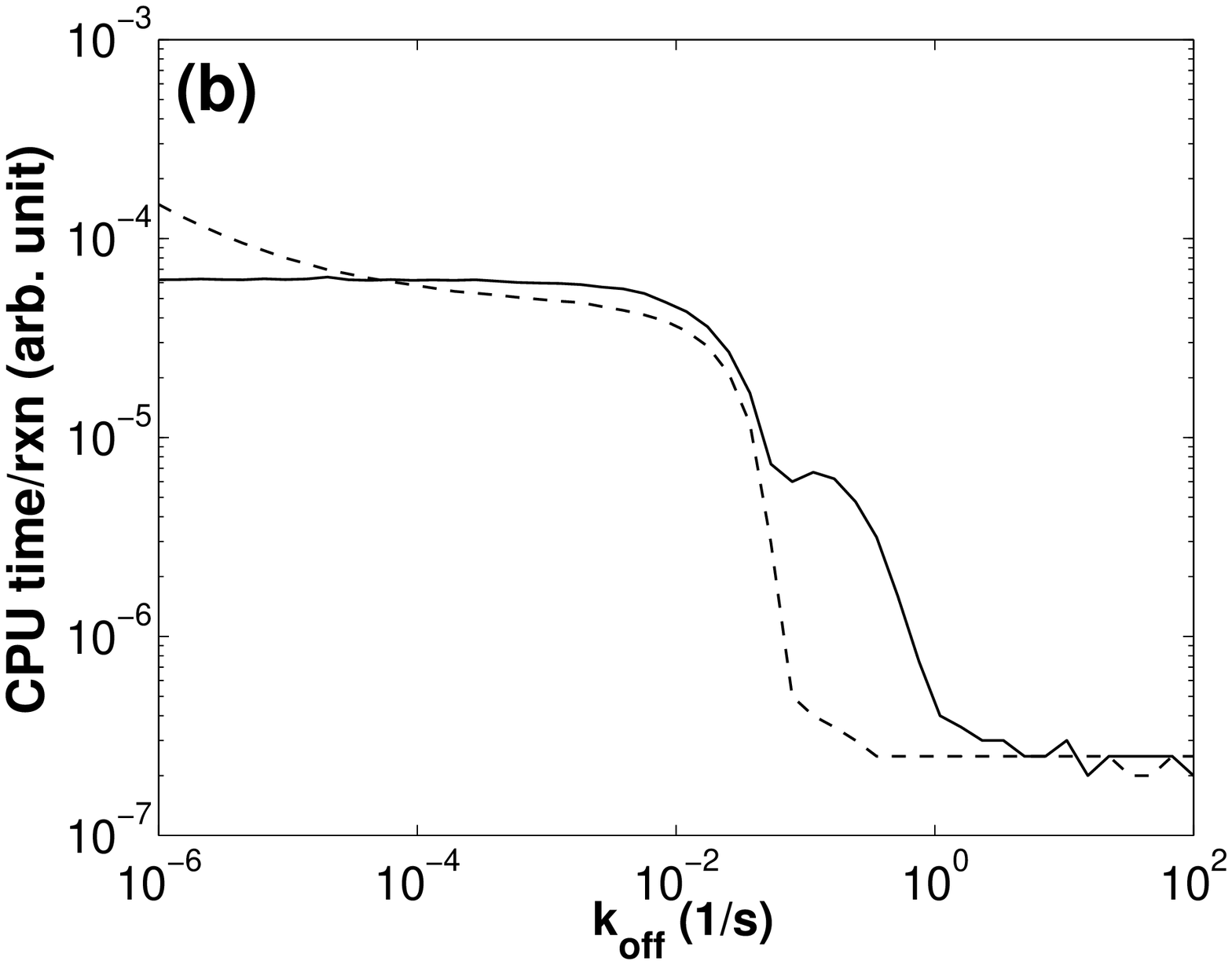} 

\includegraphics[scale=0.3]{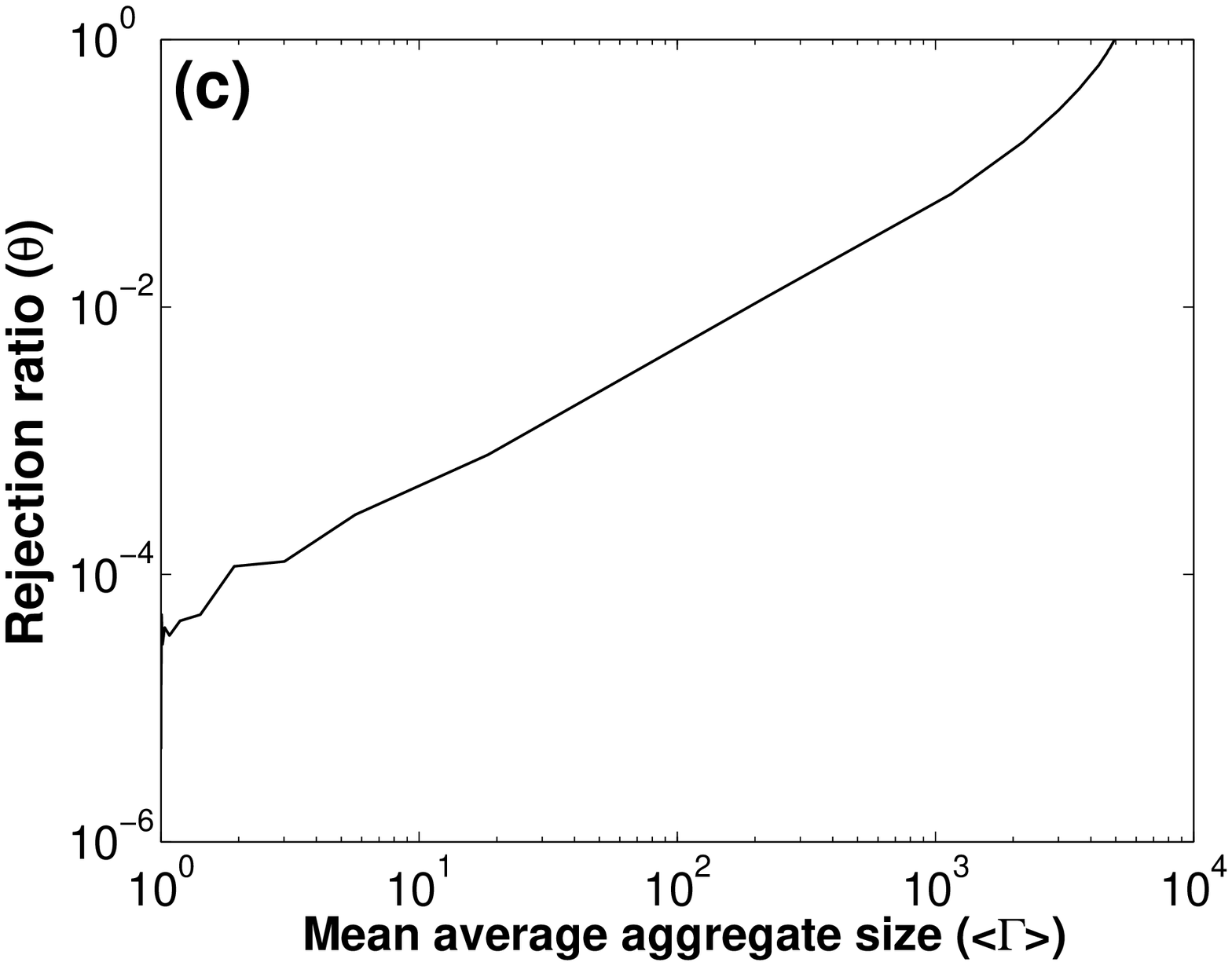}

\caption{\label{fig:perf}  Performance comparison of the rejection-free method and the rejection method. (a) The mean average aggregate size $\langle\Gamma\rangle$ (solid line) and the average number of aggregates $\langle N_A\rangle$ (dashed line). The results are normalized by the total number of receptors, $N_R$. Simulation parameters: $v_r=2, v_l=3, N_R=N_L=5000$. (b) Relationship between CPU time and $k_{\rm off}$: rejection-free method (solid line), rejection method of Yang et al.~\cite{Yang_Arxiv07} with aggregate bookkeeping (dashed line). (c) The relationship between the rejection ratio $\theta$ and the mean average aggregate size $\langle\Gamma\rangle$.  CPU times per reaction event, $\langle\Gamma\rangle$ and $\langle N_A\rangle$ are calculated by averaging over 200,000 reaction events after simulations reach equilibrium. Other parameters and initial conditions are the same as indicated in Fig.~\ref{fig:vld}.} 
\end{figure}

\begin{equation}
\Gamma=\sum_{n=1}^{N_R}np_n=\frac{1}{N_R}\sum_{n=1}^{N_R}n^2x_n \ , 
\end{equation}
where $n$ is the size of a ligand-receptor aggregate measuring the number of receptors in the aggregate and $x_n$ is the number of aggregates of size $n$. The term $p_n=nx_n/N_R$ is the fraction of receptors in aggregates of size $n$, which corresponds to the probability of finding an arbitrary receptor in an aggregate of size $n$. We note that in the above equation the number of aggregates containing single receptors, $x_1$, accounts for all receptor monomers including free receptors. Therefore, at any given time the average aggregate size $\Gamma$ takes a value in the interval $[1,N_R]$.  In practice, since the number of aggregates $N_A$ is always much less than $N_R$ (Fig.~\ref{fig:nrnc}), $\Gamma$ can be equivalently and more efficiently calculated with the following equation: 
\begin{equation}\label{eq:gamma}
\Gamma=\frac{F_R}{N_R}+\frac{1}{N_R}\sum_{j=1}^{N_A}n_j^2\ ,
\end{equation}
where $F_R$ is the number of free receptors and $n_j$ is the number of receptors in aggregate $j$. The average aggregate size $\Gamma$ can be also calculated iteratively during a simulation by accounting for the changes of aggregation caused by reaction events. To study the effect of the degree of aggregation on algorithmic efficiency by adjusting the parameters $c_{\rm tot}$ and $\beta$, we ran simulations of the TLBR model to equilibrium under a set of different values of the dissociation rate constant $k_{\rm off}$ in the range between $10^{-6}$ s$^{-1}$ and $100$ s$^{-1}$, with both $N_L$ and $N_R$ fixed at 5000. Figure~\ref{fig:perf}(a) shows a sigmoid-like relationship between the mean average aggregate size $\langle\Gamma\rangle$ and the dissociation rate constant $k_{\rm off}$. At smaller $k_{\rm off}$ (sol-gel region), $\langle\Gamma\rangle$ approaches its maximum value (close to the number of  receptors $N_R$), which indicates that a large aggregate containing most receptors exists in the system. At larger $k_{\rm off}$ (sol region),  $\langle\Gamma\rangle$ approaches its minimum value, indicating that the majority of receptors are in the form of free receptors and ligand-bound receptor monomers. The mean aggregate number $\langle N_A\rangle$ exhibits a bell-shaped curve with a maximum value of 966 near the phase transition boundary at $k_{\rm off}=0.17$. Figure~\ref{fig:perf}(b) shows the performance comparison between the rejection-free method described in this work and the rejection method of Yang et al.~\cite{Yang_Arxiv07} in terms of CPU time per reaction event, for the TLBR model with a typical set of parameters. In most of the sol region ($k_{\rm off}>1$ s$^{-1}$) and near the sol-gel region ($10^{-4}$ s$^{-1}<k_{\rm off}<10$ s$^{-1}$), the rejection-free and rejection methods match each other in performance, and the efficiency of both methods deteriorates as $\langle\Gamma\rangle$ increases. The rejection-free method is affected by searching over a near maximum number of aggregates, which is reflected in the increases of CPU time per reaction event near the boundary of the phase transition ($10^{-1}$ s$^{-1}<k_{\rm off}<1$ s$^{-1}$)  compared to that of the rejection method.  However, the rejection-free method is less sensitive to $\langle\Gamma\rangle$ and outperforms the rejection method in the highly-aggregated sol-gel region ($k_{\rm off}<10^{-4}$), where the rejection method has a rejection ratio $\theta$ greater than $99\%$ at equilibrium. As shown in Fig.~\ref{fig:perf}(c), the rejection ratio $\theta$ has a near log-linear dependence on $\Gamma$ over a wide range.

To investigate the effect of valence, we varied the number of ligand binding sites $v_l$ and kept receptor valence fixed at $v_r=2$ (Fig.~\ref{fig:perf}(b)), which can be accomplished experimentally through synthesis of multivalent ligands~\cite{mammen1998polyvalent,kiessling2000sml,posner2007trivalent}. Compared to the method of Yang et al.~\cite{Yang_Arxiv07}, the rejection-free method has better scaling of
CPU time per reaction event with the increase of $v_l$ compared to the rejection method. Except for the system with bivalent ligands, the rejection-free method is almost insensitive to changes in the number of ligand binding sites, whereas sampling by the rejection method involves large numbers of null events due to increases in intra-aggregate combinations of available ligand and receptor site pairs.

\begin{figure}
\centering
\includegraphics[scale=0.3]{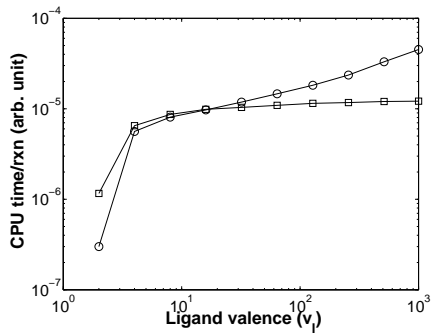}
\caption{(c) Scaling with number of binding sites on the ligand ($v_l$). For both the rejection-free (squares) and the rejection (circles) simulations, the receptor valence is fixed at $v_r=2$ and the copy numbers of receptor and ligand are $N_R=N_L=1000$. CPU times per reaction event are calculated by averaging over 200,000 reaction events after simulations reach equilibrium. Other parameters and initial conditions are the same as indicated in Fig.~\ref{fig:vld}.}
\end{figure}

\section{Discussion}
We have presented a rejection-free method for simulating biochemical reaction models specified by reaction rules. A kinetic Monte Carlo procedure is applied to sample the rule list, identify reactant proteins, and update protein states. In this procedure, all chemical species are formed dynamically. For this reason, our method has a computational cost independent of reaction network size.  The implementation described here is a rejection-free method in which every Monte
Carlo step changes the state of the system. The method accounts for the exact
rates of reaction rules and probability distributions of protein sites, including rules that require evaluation of non-local state information. In a more general and straightforward implementation, selecting candidate reactant sites based on these probability distributions incurs linear-time cost per reaction event scaled by the number of sites. In contrast, the rejection-free implementation reported here in fact scales with the number of existing chemical species. In the multivalent ligand-receptor interaction model, the number of chemical species during a simulation corresponds to the number of ligand-receptor aggregates, $N_A$. Although the rejection-free method has a higher cost per Monte Carlo step because of searching candidate sites and extra bookkeeping required for calculating probabilities for sites, the method outperforms the rejection method when the rejection ratio $\theta$ approaches unity. Our method is closely related to the direct simulation Monte Carlo (DSMC) method developed to simulate coagulation processes where irreversible particle aggregation is considered~\cite{kruis2000direct}.

A comparison between rejection-free and rejection methods was briefly discussed by Yang et al.~\cite{Yang_Arxiv07}. The rejection-free algorithm reported by Yang et al.~\cite{Yang_Arxiv07} takes a strategy of searching for a ligand-receptor site pair for cross-linking, which is less efficient for simulating formations of large ligand-receptor clusters in comparison to the algorithm presented in this report. This earlier algorithm maintains reaction probabilities of all sites for searching, which has a cost of CPU time per reaction event scaling with the number of molecules. In contrast, the current method searches over an aggregate list that is compact compared to the full list of molecule sites. For example, as shown in Fig.~\ref{fig:perf}(a), the maximum number of aggregates on average (observed around $k_{\rm off}=0.1$ s$^{-1}$) was less than one fourth of the total number of receptors ($N_R=5000$). At other values of $k_{\rm off}$, $\langle N_A\rangle$ is much less than $N_R$ (Fig~\ref{fig:perf}(a)). As shown in Fig.~\ref{fig:nrnc}, at equilibrium $\langle N_A\rangle$ has a moderate (sublinear) dependence on the number of molecules.

Both rejection and rejection-free methods require explicitly tracking connectivity between sites. This feature erodes the efficiency of simulation. In simulating the multivalent ligand-receptor binding model, to process an aggregate dissociation, at least one unweighted traversal of an aggregate subgraph is necessary (with no prior information about which subaggregate is smaller). This presents a major bottleneck to simulating systems in the sol-gel regime because a graph traversal has an order of cost proportional to the size of the aggregate. In the sol-gel regime, most graph traversals will happen on the giant aggregate that has a size close to that of the entire system. Experiments can provide some information about the composition of a protein aggregate but  the intra-aggregate (or intra-multiprotein complex) topology usually cannot be resolved. To compare with data, it may be possible to simulate a system without tracking the aggregate topology if prediction of connectivity information (e.g., distribution of aggregate topologies for an aggregate size) is not of interest.

In summary, our results suggest that a rejection-free kinetic Monte Carlo approach to simulation of rule-based models has appeal for simulating aggregation processes, which are common in many biochemical systems. Although our implementation in this study is problem-specific for multivalent ligand-receptor interactions, our results suggest that a generalized implementation of rejection-free procedures along the lines presented here into existing rule-based modeling software may be beneficial. 

\section*{Acknowledgments}
We thank J. R. Faeder, M. I. Monine and Q. Chang for helpful discussions. This
work was supported by NIH grants GM076570, GM085273 and RR18754, DOE contract
DE-AC52--06NA25396, and NSFC grant 30870477.   We thank the Center for Nonlinear Studies for support that enabled J.Y. to visit Los Alamos. 

\bibliographystyle{elsarticle-num-names}
\bibliography{Yang_JCP_2008}

\end{document}